# Evaporation of Colloidal Suspension Droplets over Oil-infused Slippery Surfaces

Sumita Sahoo and Rabibrata Mukherjee[*]

Instability and Soft Patterning Laboratory, Department of Chemical Engineering, Indian Institute of Technology Kharagpur, Kharagpur, WB, INDIA 721302

**E-mail id:** rabibrata@che.iitkgp.ac.in

**Abstract**

Evaporative drying of colloidal droplets with different concentrations ($C_i$) on liquid infused slippery substrate (LISS) surfaces fabricated over biomimetically patterned sticky hydrophobic surfaces is reported here. These slippery surfaces remain stable under the aqueous droplets. The effective thickness of the infused oil layer ($h_E$) was varied and its effect on the evaporation dynamics, wetting ridge formation and the final deposition pattern were studied using an optical microscope as well as a contact angle goniometer. The results were contrasted to those obtained over bare (oil-free) flat and patterned sticky surfaces. Unlike strong pinning and consequent peripheral ring deposit invariably observed in evaporation of a colloidal droplet over a solid surface (i.e. bare flat Sylgard 184 without oil), over a LISS, the dynamics, including the extent of pinning and the final deposit morphology, were found to be strongly dependant on $h_E$. Due to the presence of the underlying patterns, the extent of pinning increased with a reduction in $h_E$. Prolonged evaporation over a LISS leads to a uniform deposition due to gravity-induced settling of the colloids followed by capillarity-mediated rearrangement of the particles. Colloidal assembly at the droplet free surface and the droplet-wetting ridge interface leads to a thinner encapsulating oil layer and faster drainage of the wetting ridge resulting in a higher evaporation rate before pinning of the suspension droplets over all LISS surfaces.

# 1. Introduction

Evaporative drying of a solution or colloidal suspension droplet over a rigid substrate is a ubiquitous problem that has extreme significance in several areas such as inkjet printing,[1] functional coating,[2] biosensor and diagnostics,[3] forensic science[4-5] SERS-related plasmonic applications,[6] and so on. As shown by Picknett and Bexon, depending on the nature of the substrate as well as the extent of pinning due to substrate heterogeneity and strength of adhesion, a droplet can evaporate either in: 1) constant contact angle (CCA) mode; or 2) constant contact radius (CCR) mode.[7] However, on most practical surfaces a mixed mode involving a transition from an initial CCA to CCR mode is generally observed. A pinned contact line engenders a ring-like deposit at the periphery of the droplet, where due to the effect of curvature the rate of evaporation is maximum and in order to compensate for the rapid solvent loss from there, a radially outward advective flow sets in within the droplet, due to a combination of Solutal Marangoni effect, thermo-capillary effect and osmotic pressure gradient. This leads to the migration of the solute particles/ colloids towards the pinned periphery, resulting in the formation of the concentric "coffee ring patterns" which was popularized by Deegan,[8-14] and has received significant research attention ever since.

On the other hand, there is rather limited literature on drying of a droplet over a soft deformable surface, where the vertical component of surface tension acting along the three-phase contact line (TPCL) causes peripheral deformation.[15-17] Of particular interest is the evaporative drying of a colloidal droplet over artificially fabricated slippery surfaces, design and development of which is strongly inspired by naturally occurring slippery leaves such as 'Nepenthes alata' or pitcher plant.[55] Fabrication of a variety of such slippery surfaces has been reported in the last few years, which depending on the exact technique of fabrication, has been termed lubricant-impregnated surfaces (LIS)[18-22] or slippery liquid-infused porous surfaces (SLIPS)[23-28]. Such surfaces have been used extensively in the development of self-cleaning,[23] and anti-biofouling surfaces,[25,29] slippery papers,[30] droplet-based microfluidics with directional mobility control,[31-33] synthesis of supra-particle assembly,[15-16] and so on. On such a surface, a liquid droplet slides rather easily without getting pinned. However, in order to satisfy the Neumann force triangle based on the balance of $\gamma_{wa}$ (droplet – air surface tension), $\gamma_{wo}$ (droplet – oil interfacial tension), and $\gamma_{oa}$ (oil – air surface tension,) acting at the TPCL (refer to figure 1), a unique annular lubrication ridge, better known as the wetting ridge is seen to form surrounding the droplet periphery.[18] Further, depending on the spreading coefficient of the infusing oil over the droplet, the lubricating ridge often climbs over the droplet free surface and completely cloaks it,[18] resulting in a reduced rate of evaporation.[27] In certain cases, it has been observed that as the lubricating oil rises around the droplet periphery during wetting ridge formation, the oil layer below the droplet may undergo spontaneous rupture and dewetting due to reduced local thickness of the film, leading to loss of slipperiness of the surface. Recently, we have developed a slippery surface by dispensing high-viscosity silicone oil over a topographically patterned sticky cross-linked hydrophobic Sylgard 184 surface obtained by biomimetic replication of Rose petals. The presence of hierarchical patterns on such surfaces anchors the oil layer and prevents its rupture in case of local thinning of the oil layer or due to gravity-driven drainage of the slip-enhancing oil layer, thereby retaining the slipperiness of the surface. Further, the extent of slipperiness over such a surface can be modulated by varying the oil layer thickness ($h_E$), which in turn significantly influences the evaporation time as well as the wetting ridge dynamics during the evaporation of a pure water droplet.

In this article, we explore the evaporative drying of an aqueous colloidal droplet over such slippery surfaces fabricated by infusing silicone oil over a sticky hydrophobic base substrate. Interestingly, there are very few published articles on evaporative drying of a colloidal or solution droplet over a slippery surface.[15-17,21,34] Chaniel et al. reported evaporation of a salt solution droplet over a slippery, porous polymer substrate and observed the formation of a solute-rich peripheral region as the solvent was seen to preferentially migrate inwards due to a combination of both Solutal and thermal Marangoni flows.[34] In contrast, a more uniform deposition was obtained by Mcbride et al. in drying of an almost identical system over a slippery surface fabricated by liquid impregnation over a textured substrate, where the final deposition pattern was found to be largely independent of the precise geometry of the substrate patterns.[21] Gao et al. studied the evaporation of aqueous colloidal suspensions microdroplets over oil-coated flat substrates and observed the formation of supra-particle assemblies when the thickness of the oil layer was above a critical value.[15] Further, they showed that the particles flow towards the free water-air interface beyond the wetting ridge as the rate of evaporation is higher there and eventually results in a uniform colloidal deposition after the late-stage collapse of the skin layer.[15] Li et al. showed that the deposition morphology of polymer microspheres over an oil-wetted slippery surface could be tuned by the addition of surfactant, where contact line pinning was achieved due to enhanced surfactant concentration and not due to an increase in the particle concentration.[17] Similarly, Kim et al. reported the formation of supra-particle assemblies of different shapes like hemisphere, convex lens, flat discs and so on by evaporative drying of a surfactant-laden colloidal dispersion droplet over a slippery surface by modulating the apparent contact angle as well as the contact line friction in presence of surfactants.[16]

It becomes evident from the preceding discussion that there have been no studies on how the deposition morphology is influenced by the infusing oil layer thickness ($h_E$), when a colloidal dispersion droplet of different colloidal concentration ($C_i$) evaporates over a slippery surface. We show how the droplet diameter ($d_D$), droplet height ($h_D$), apparent contact angle ($\theta_{app}$) as well as the height ($h_R$) and the width ($w_R$) of the wetting ridge evolves with time ($t_E$) during evaporation of colloidal droplets on different LISS surfaces having different $h_E$ and correlate them to the final deposition patterns. We could perform this study, particularly at low oil layer thickness, because of the stability imparted by the oil layer by the underlying substrate patterns, preventing any possible loss of slipperiness. In fact, this study allowed us to understand how the evaporation dynamics as well as the deposition morphology changes as a function of the ratio of $h_E$ and the feature height ($h_P$), which strongly influences the extent of slipperiness. Consequently, we observe unique phenomena like the formation of multiple overlapping wetting ridges and so on, under certain specific conditions. Based on our experiments, we conclude how the final deposit patterns over different LISS surfaces strongly depend on the dominant mechanism for colloidal assembly at the different interfaces, that is at the droplet free surface, droplet base and droplet-wetting ridge interface, rather than the value of $C_i$.

2. **Material and Methods**

LISS surfaces were fabricated by spin coating silicone oil (Sigma-Aldrich; 1000 cSt) onto a soft lithographically fabricated positive replica of a rose petal on crosslinked Sylgard 184 (a two-part thermo-curable elastomer, Dow Corning, USA) films fabricated by a sequential double replication process.[37] The replicated patterns are nearly identical to the patterns on an actual rose petal, which are hierarchical in nature comprising conical primary micro pillars decorated with secondary nano folds.[37] The average feature height of the pillars ($h_P$) on the actual rose petals as well as on the replicas was found to be $h_P \approx 7.0 \pm 0.4$ μm, indicating perfect replication with nearly no loss of fidelity. The equilibrium contact angle ($\theta_E$) measured using a contact angle goniometer was found to be $\theta_E \approx 137.8° \pm 3.0°$ on an actual rose petal and $\theta_E \approx 135.5° \pm 3.2°$ on the replica. The replica was also sticky, like the original rose petal as a water drop was seen to remain completely stuck, even when the configuration of the drop was changed from sessile to pendant by tilting the surface by 180°.[38]

For fabrication of slippery surfaces, square pieces (2.5×2.5 cm$^2$) of the patterned Sylgard 184 substrates were spin-coated with 0.5 ml silicone oil using a spin coater (Apex Instruments, India). The effective oil layer thickness ($h_E$) was controlled by simultaneously varying the RPM and duration of spinning, details of which are given in table 1. After spin coating, the coated films were kept in a horizontal configuration for 30 min at room temperature, followed by heating at 100°C for 30 min in a hot air oven for the uniform spreading of the oil layer.[37] It is well understood that a flat film is not obtained over a topographically patterned surface after spin coating, as the local thickness of the coated film is thinner over the protrusions and thicker over the recessed zones.[40-41] Thickness of such films can be quantified by "effective film thickness" ($h_E$), representing the thickness of a film coated on a flat substrate of the same material under identical spin-coating conditions.[37,42-44] In this work, we varied the $h_E$ of the oil layer between ≈ 10 μm and 100 μm by varying the spin speed and time as shown in table 1.

*Table 1: Characteristics of different LISS surfaces*

| Substrate | Silicone oil coating condition | | Surface Characterization | | | Oil layer thickness | | Comments |
| --- | --- | --- | --- | --- | --- | --- | --- | --- |
| | Spin RPM | Spin time (sec) | $\theta_{app,i}$ (°) | CAH (°) | Sliding angle (°) | $h_E$ (μm) | $\frac{h_E}{h_P}$ | |
| LISS 1 | 500 | 30 | 85.5 ± 3.5 | 0.0 ± 0.2 | 0.1 | 99.5 ± 2.3 | 13.8 | Oil-flooded |
| LISS 2 | 1500 | 120 | 95.8 ± 4.0 | 2.2 ± 0.2 | 1.6 | 15.7 ± 1.0 | 2.1 | Properly oil-coated |
| LISS 3 | 3500 | 120 | 102.0 ± 4.5 | 3.7 ± 0.4 | 2.1 | 12.9 ± 1.5 | 1.8 | |
| LISS 4 | 6000 | 120 | 110.0 ± 3.0 | 4.0 ± 0.5 | 3.2 | 8.1 ± 1.8 | 1.1 | Oil-starved |
| Rose Petal replicated Sylgard 184 | - | - | 135.5 ± 3.5 | - | - | 0.0 | - | Bare sticky surface |

Table 1 summarises various aspects of the fabricated LISS surfaces having different $h_E$. In the lines of the terms used by Xiong et al.,[61] we categorise the fabricated LISS samples into three categories depending on the values of $h_E/h_P$: (a) LISS 1 ($h_E$ ≈ 99.5 μm, $h_E/h_P$ >> 1) identified as "oil-flooded" LISS; (b) LISS 2 and 3 ($h_E$ ≈ 15.7 μm and 12.9 μm respectively, $h_E/h_P$ ≈ 2.0 to 1.5), identified as "properly oil-coated" LISS; and (c) LISS 4 ($h_E$ ≈ 8.1 μm, $h_E/h_P$ ≈ 1), termed as "oil-starved" LISS. The ratio ($h_E/h_P$) qualitatively provides an idea about the relative thickness of the oil layer with respect to the feature height ($h_P$) of the micropillars on the patterned substrates. Aqueous droplet over these LISS surfaces acquires a 'slippery Wenzel wetting state' instead of a 'hemi wicked Cassie-Baxter wetting state', as shown in our previous work, due to the stable oil layer over the conical micropillars due to the secondary sticky nano-folds.[37] It can be seen in table 1 that despite significant decrease in $h_E$, there is only a marginal increase in CAH from ≈ 0.1° on a LISS 1 to CAH ≈ 4.0° on a LISS 4 surface, as can be seen in table 1, clearly indicating that all the LISS samples retain slipperiness. The extremely low values of CAH are characteristic of slippery surfaces as the water layer essentially slides over a high-viscosity silicone oil layer. In our context, it is extremely important to highlight that the nano-folds on the patterned base substrate which are known to impart stickiness actually anchor the oil layer, thereby preventing its flow. However, these patterns do not restrict the sliding of a water droplet in any way as the droplet base never come in direct contact with these folds. Interestingly, the initial apparent contact angle ($\theta_{app,i}$) changes quite significantly from $\theta_{app,i,LISS1}$ ≈ 85.5° to $\theta_{app,i,LISS4}$ ≈ 110.0°, arguably as the effect of the substrate patterns become more pronounced with the reduction in $h_E$, alongside the inherent hydrophobicity of silicone oil.

Droplets of aqueous suspension of colloidal polystyrene (PS, mean diameter ≈ 3 μm; Sigma Aldrich) in distilled water with initial colloid concentration ($C_i$) varying between 0.015% and 1.25% (w/v) were dispensed on the different LISS surfaces. The colloidal suspension was stabilized using an ultrasonic cleaner (Microclean-103) for 15 mins before dispensing. The droplet volume was kept constant at 10 μl in all cases. The evaporation dynamics and internal flows were observed in situ under a stereo-zoom (optical) microscope (Leica Microsystems Ltd. DMC2900) attached to a digital CCD camera and a "Boom" stand. The final colloidal deposition patterns were examined using a compound optical microscope (Leica DM 2500). All apparent contact angle ($\theta_{app}$), as well as contact angle hysteresis (CAH) related measurements, were performed using a contact angle goniometer (Apex Instruments Co. Pvt. Ltd., India, Model: ACAMHSC 03). CAH is measured by dispensing 10 μl water droplets on each liquid-infused slippery surface while the sample stage is tilted at a rate of 0.5°/min. All the experiments were performed at 23°C and 27% relative humidity (RH).

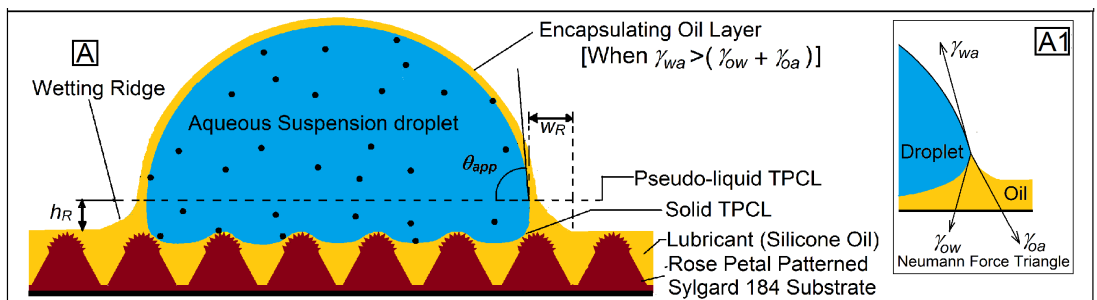

*Figure 1: Schematic of an aqueous colloidal suspension droplet dispensed over a LISS surface. Inset A1 shows the balance of the interfacial surface tension forces in a Neumann force triangle.*

It has already been mentioned that one unique aspect of dispensing a droplet over a liquid-infused slippery surface (LISS) is the formation of the wetting ridge, due to the vertical component of interfacial tensions acting at the TPCL. To facilitate subsequent discussion, this is schematically explained in figure 1, which clearly indicates the height ($h_R$) and width ($w_R$) of the wetting ridge as well as the apparent contact angle ($\theta_{app}$) the droplet makes with respect to the horizontal plane at the pseudo-liquid TPCL.[18,37] The wetting ridge dimensions ($h_R$ and $w_R$) were measured by post-processing of the goniometer images by using the software "image J".

3. **Results and discussions**

   *3.1. Effect of variation of $h_E$ on evaporation dynamics and final deposit patterns*

We first show how the evolution sequence, as well as the final deposition morphology in evaporative drying of a colloidal droplet (with $C_i \approx 0.075\%$), differs over a slippery surface (silicone oil coated Rose petal replica, series C of figure 2) in contrast to that observed over a flat hydrophobic surface (flat cross-linked PDMS surface, series A of figure 2) as well as a patterned sticky hydrophobic surface (positive replica of Rose petal, series B of figure 2). The aqueous colloidal suspension droplet gets fully encapsulated with a layer of silicone oil as the spreading coefficient of silicone oil over water in the air, $S_{ow(a)} \approx 7.7$ mN/m (positive). The formation of the cloaking layer can be verified from the ripples that can be seen on the droplet surface.[37] Such ripples were totally absent in droplets dispensed over flat and patterned bare Sylgard 184 surfaces. The encapsulation layer, along with the wetting ridge reduces the rate of evaporation of the droplet, which results in a much longer time required for complete evaporation ($t_F \approx 191$ min, figure 2 C6) over a slippery surface, as compared to $t_F \approx 51$ min (figure 2 A6) and $t_F \approx 62$ min (figure 2 B6) over a flat and patterned Sylgard 184 surface respectively. These images also show that the final deposit morphology including its size and nature of the deposit is significantly different on the three types of surfaces.

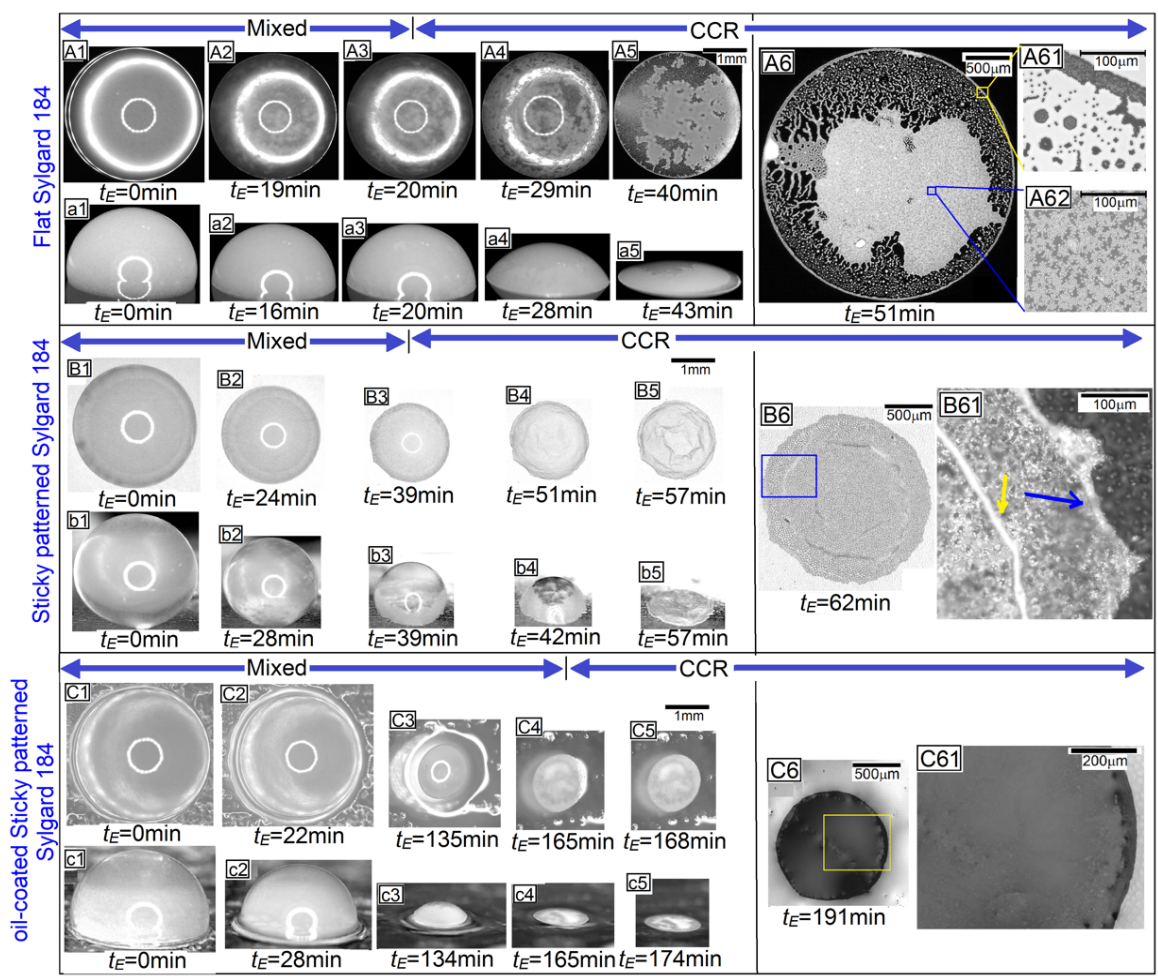

*Figure 2: Optical micrographs of an aqueous suspension droplet having $C_i \approx 0.075\%$ with progressive evaporation time ($t_E$) over (A1-A6) flat Sylgard 184 substrate; (B1-B6)Rose petal patterned sticky hydrophobic Sylgard 184 substrate and (C1-C6) oil-coated sticky patterned Sylgard 184 substrate (LISS 2 surface). Insets A61-A62, B61 and C61 show magnified optical micrographs of the final deposit pattern.*

Over a flat substrate, the droplet periphery gets pinned within $t_E \approx 20$ min, which leads to the formation of a final deposit that has $d_F \approx 2.96$ mm. The circular deposition zone is almost entirely covered with particles (inset A62, figure 2) and surrounded by a clear peripheral coffee ring (inset A61, figure 2). Between these two zones, there exists a region with scattered particle deposition. On the other hand, over a sticky hydrophobic surface, the contact line continues to retract longer, for $t_E \approx 39$ min, leading to a significant reduction in the droplet diameter. This is extremely interesting and counter-intuitive, as due to its sticky nature, the contact line of a pure water droplet dispensed over such a surface gets pinned almost immediately after deposition and evaporates entirely in the CCR mode.[37] The stickiness of such substrates is a result of the 'Cassie impregnated Wenzel state' due to the penetration of water into the nano-folds and the micro-pillars.[38] However, when a colloidal dispersion is dispensed over such a hierarchically patterned substrate, we feel that some particles probably accumulate towards the periphery and temporarily deposit over the nano folds, leading to an in-situ loss in their anchoring capability. This in return, allows the contact line to retract over the micropillars while leaving behind a few colloids deposited in between the pillars. This is a completely new and novel observation, though beyond the scope of the present manuscript and will be taken up in detail separately. The final deposit comprises of a fully particle-covered central region with a clear signature of multi-layer deposition with surface wrinkles which are marked with a yellow arrow in inset B61 of Figure 2 surrounded by a thin peripheral ring marked with a blue arrow in the same figure.

In contrast to both the cases, over LISS 2 surface, the suspension droplet retracts freely for most of its lifetime and gets pinned only at a later stage after $t_E \approx 165$ min forming a comparatively smaller final deposit zone as can be seen in frame C6 of figure 2. During this prolonged evaporation period, most of the colloids settle randomly at the droplet base due to gravity resulting in an almost uniform deposit. From the optical microscope images, it appears that the deposited particles penetrate into the oil layer in the final deposit (figure 2 C61), though we could not confirm it more accurately as FESEM imaging was not possible.

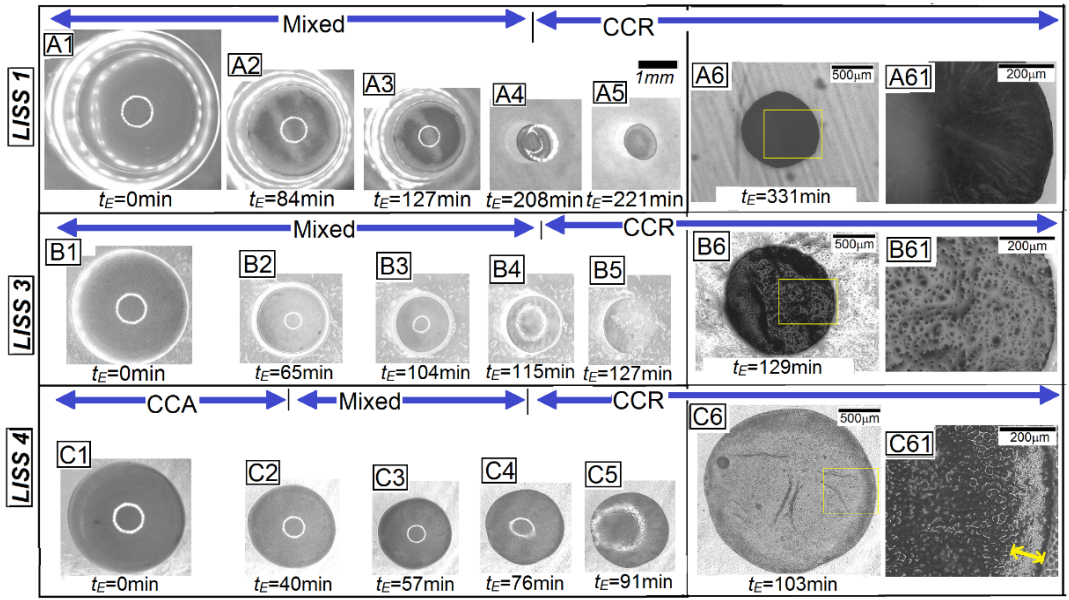

*Figure 3: Optical micrographs of an aqueous suspension droplet having $C_i \approx 0.075\%$ with progressive evaporation time ($t_E$) over different LISS surfaces: (A1-A6) LISS 1; (B1-B6) LISS 3 and (C1-C6) LISS 4 surfaces. Insets A61, B61 and C61 show magnified optical micrographs of the final deposit near the periphery.*

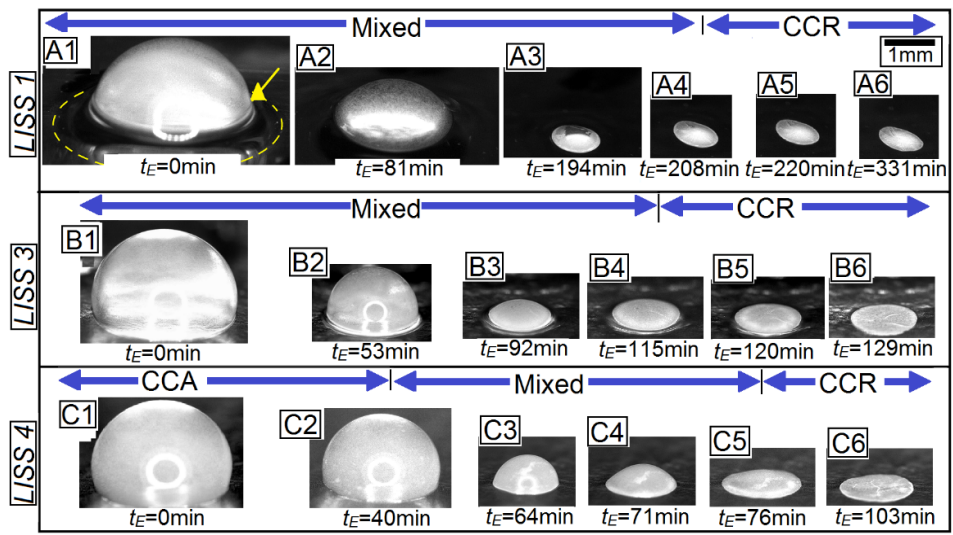

*Figure 4: Side view optical micrographs showing the temporal evolution of an evaporating aqueous suspension droplet having $C_i \approx 0.075\%$ over: (A1-A6) LISS 1; (B1-B6) LISS 3 and (C1-C6) LISS 4 surfaces.*

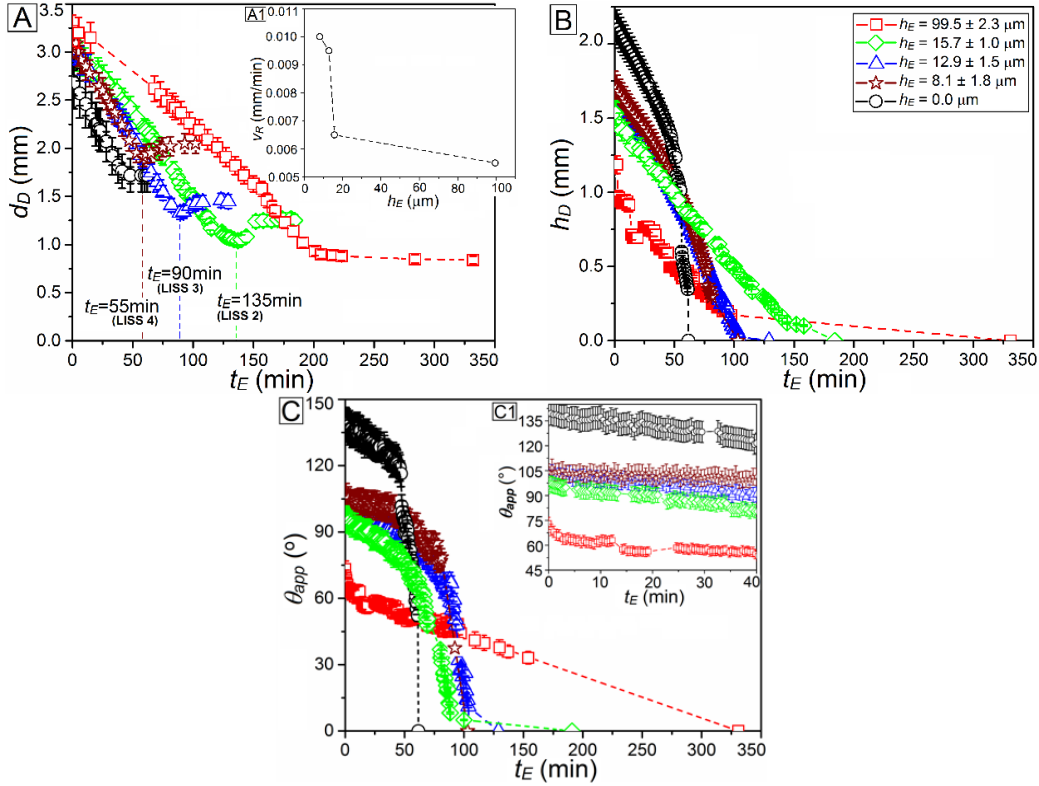

*Figure 5: Variation of the droplet (A) diameter ($d_D$), (B) height ($h_D$) and (C) apparent contact angle ($\theta_{app}$) with $t_E$ for an aqueous suspension droplet of $C_i \approx 0.075\%$ over different LISS surfaces with varying $h_E$. Inset A1 shows the variation of average contact line retraction velocity ($v_R$) as a function of $h_E$. Inset C1 shows a magnified version of the initial part of the plot (C) up to $t_E \approx 40$ min.*

Once we have shown how the deposition pattern as well as $t_F$ is significantly different on a LISS surface in comparison to an oil-free surface, we now move on to explore how $h_E$ influences the evaporation dynamics as well as final deposit patterns. For this purpose, the evaporation of a colloidal suspension droplet with $C_i \approx 0.075\%$ on different LISS surfaces with varying $h_E$ was studied. The image frames in figure 3 and 4 show the top view and side view of the evaporating droplet on different LISS surfaces at different $t_E$. Further, frames A, B and C of figure 5 show how the droplet diameter ($d_D$), droplet height ($h_D$) and the apparent contact angle ($\theta_{app}$) progressively change with $t_E$ on the different LISS surfaces. Similar to what has been observed with a pure water droplet, we observe that with a gradual reduction in $h_E$, $\theta_{app,i}$ of the dispensed droplet increases. This is arguably due to the fact that as $h_E$ reduces, the top surface of the silicone oil does not remain flat anymore and becomes undulating along the contours of the substrate patterns.[54] Consequently, the roughness on the oil layer leads to an increase in $\theta_{app,i}$. It is also observed that on each LISS surface, $\theta_{app,i}$ is slightly lower for a colloidal suspension droplet as compared to a pure water droplet of the same volume,[37] arguably due to the presence of the PS colloids which has lower surface tension ($\gamma_{Pa} \approx 42$ mN/m)[45] as compared to pure water ($\gamma_{wa} \approx 72$ mN/m).[38]

It is also seen that the droplets initially evaporate in the mixed mode with a simultaneous variation in $d_D$ and $\theta_{app}$, followed by CCR mode when $d_D$ remains constant after the contact line gets pinned. (figure 5A) over all the LISS surfaces except over LISS 4, where it initially evaporates in CCA mode ($\theta_{app}$ remains constant as shown in inset C1 of figure 5C) followed by an intermediate mixed mode (between $t_E \approx 40$ min and $t_E \approx 76$ min), and finally in the CCR mode beyond $t_E \approx 76$ min. Further, on all the LISS surfaces except LISS 1, there is an intermediate increase in $d_D$ at a later stage (marked in figure 5A), and discussed in detail later. Further, it is seen in figure 5A that on each of the LISS surfaces, the retracting velocity of the contact line ($v_R = \frac{1}{2}\frac{\Delta d_D}{\Delta t_E}$) remains nearly constant for a significant duration of time, till the droplet gets pinned.

From figure 3 and 4, we observe that $t_F$ increases and the diameter of the final deposit ($d_F$) decreases with an increase in $h_E$. The increase in $t_F$ with $h_E$ is directly attributed to the higher availability of silicone oil, which results in higher and thicker wetting ridges. Due to the same reason, pinning is deferred on LISS surfaces with higher $h_E$, allowing the contact line to retract freely, leading to a final deposition with reduced $d_F$. For example, over a LISS 1 surface, $d_F \approx 0.85 \pm 0.05$ mm (figure 3A6). In contrast, $d_F \approx 1.35 \pm 0.05$ mm and $1.43 \pm 0.05$ mm over LISS 2 and LISS 3 surfaces respectively, as can be seen from figure 2C6 and figure 3B6. On the other hand, over a LISS 4 surface, an almost uniform colloid deposit with $d_F \approx 2.02 \pm 0.05$ mm with a thin peripheral ring is observed (figure 3 C6). It is also observed that the final colloidal deposit is partially submerged under silicone oil over all the LISS surfaces. This conclusion is drawn from the higher magnification images presented in frames A61, B61 and C61 of figure 3.

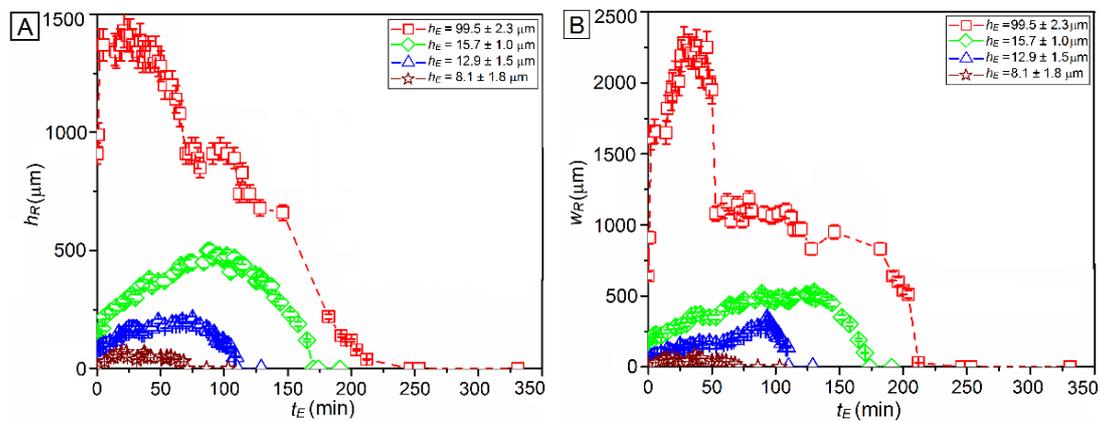

*Figure 6: Temporal evolution of wetting ridge (A) height ($h_R$) and (B) width ($w_R$) for an evaporating aqueous suspension droplet having $C_i \approx 0.075\%$ on different LISS surfaces with varying $h_E$.*

Next, we investigate how the height ($h_R$) and the width ($w_R$) of the wetting ridge vary with $t_E$ on different LISS surfaces. We observe that on all the LISS surfaces except LISS 4 surface, both $h_R$ and $w_R$ initially grow, and attain a maximum $h_{R-max}$ and $w_{R-max}$. Beyond this stage, the wetting ridge gradually drains. It is observed that $h_{R-max}$ is highest ($\approx$ 1500 μm) over a LISS 1 surface (figure 6A). In contrast, the wetting ridge shows limited growth over LISS 2 and LISS 3 surfaces and is almost absent over a LISS 4 surface. With a gradual reduction in $h_E$, on one hand less amount of oil is available for ridge formation and on the other hand, the nano-folds present on the patterned base substrate impart higher anchoring to the oil layer. Because of these two combined effects, $h_{R-max}$ gradually reduces with the reduction in $h_E$. Similar variation is also observed in $w_{R-max}$ over different LISS surfaces. For example, over a LISS 1 surface, the reduction in $w_R$ after achieving a maximum ($w_{R-max} \approx$ 2300 μm) at $t_E \approx$ 26 min is non-monotonic as it remains nearly constant at $w_R \approx$ 1000 μm in between $t_E \approx$ 55 min and $t_E \approx$ 147 min (figure 6 B). This happens due to the highly viscous nature of the thick oil layer. In contrast, a much thinner wetting ridge is initially observed due to a combination of lower availability of oil (for lower $h_E$) as well as strong anchoring imparted by the sticky underlying surface to the oil layer over LISS 2 and LISS 3 surfaces. And lastly, on an "oil-starved" LISS 4 surface, the wetting ridge is almost non-existent initially and a faint transient ridge is observed after $t_E \approx 20$ min before the ridge starts to drain as shown in figure 6. It is already known that gradual reduction in $h_R$ is associated with the drainage of the encapsulating oil into the wetting ridge due to the capillary pressure arising out of droplet curvature that renders the base of the wetting ridge with a smaller radius of curvature a low-pressure region.[47] Finally, the interrelation between $h_{R-max}$, $w_{R-max}$ and $h_E$ can be explained by shear stress at both oil–water and oil–air interface during evaporation which can be expressed as $\tau_i = \mu_o V_i'/h_E$;[47] where $\mu_o$ is oil viscosity and $V_i'$ is the interfacial velocity of oil. For LISS surfaces with higher $h_E$, the shear stress at both interfaces will be lower, leading to a faster flow of oil. This effect, along with the availability of more oil is responsible for higher $h_{R-max}$ being attained rapidly in the case of a LISS 1 surface. It is important to point out that the qualitative nature of variation of $h_R$ and $w_R$ with $t_E$ on different LISS surfaces during the drying of a colloidal droplet is rather similar to that observed during the evaporation of a pure water droplet. However, there are some subtle differences between the two cases. For all the LISS surfaces, evaporation of a colloidal droplet continues for much longer after the disappearance of the wetting ridges, in comparison to a pure water droplet over LISS surfaces, where the flattening of the wetting ridge and completion of drop evaporation almost coincides. The early drainage of the wetting ridge during evaporation of a colloidal droplet is attributed to a reduction in the interfacial tension at the droplet–wetting ridge interface, due to the accumulation of colloidal particles along the ridge. As the PS colloid has lower surface tension than water, the interfacial tension reduces and consequently, the vertical component of interfacial tension which is responsible for the wetting ridge formation fails to satisfy the balance of interfacial forces acting along the TPCL and as a result, the wetting ridge starts to drain.

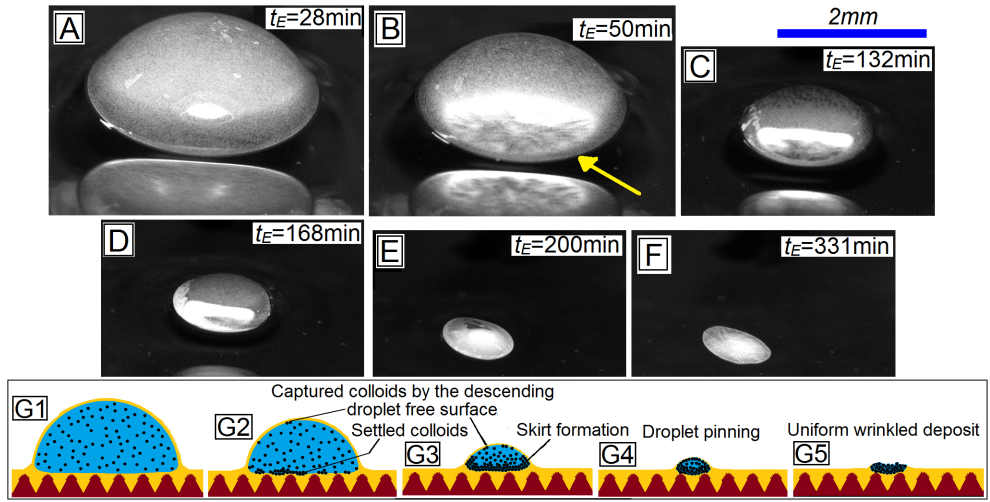

*Figure 7: Side view optical micrographs showing colloidal assembly at the droplet base as well as at the droplet free surface during evaporation of an aqueous suspension droplet having $C_i \approx 0.075\%$ over a LISS 1 surface. Frames G1-G5 show the evolution of colloidal self-assembly schematically.*

Apart from the variation in the final deposition morphology, we observe that $h_E$ also influences the deposition mechanism of the colloids on different LISS surfaces, which is shown in figure 7 and 8. Frames A-F of figure 7 show the deposition sequence when a colloidal droplet with $C_i \approx 0.075\%$ is dispensed over a LISS 1 surface. Initially, no significant colloidal deposition is observed either at the base or at the free surface of the droplet (figure 7A). After $t_E \approx 50$ min, we observe gradual deposition of the colloidal particles at the droplet base (figure 7B) while exhibiting some radially outward transient fractal-like patterns which subsequently disappear with the deposition of more number of particles with progressive $t_E$. Till this stage, the retracting TPCL does not show any pinning, and the free surface of the droplet remains almost particle free. After $t_E \approx 132$ min, a particle-rich skirt-like zone is seen to rise near the TPCL. The formation and growth of this particle-rich zone are attributed to multiple factors. Firstly, we have already discussed that the droplet base-oil interface gets covered with particles. So, when the TPCL retracts further, the particle-covered droplet base simply folds upward along the inner surface contours of the wetting ridge as schematically shown in figure 7G.

At the droplet base, the colloids get reorganised with progressive retraction of the contact line to form a fractal-like pattern followed by complete coverage of the droplet base by a compact colloid layer. This phenomenon is similar to what was observed by Dumazer et al. during the capillary bulldozing of sedimented particles in a milli-fluidic tube, where either a 'viscous displacement regime' or a 'frictional displacement regime' exists depending on the fluid velocity, particle concentration and size.[56-57] In our system, initially, a 'viscous displacement regime' exists when the area fraction of deposited colloids covering the droplet base is small, resulting in a radial displacement of the colloids with the retracting contact line. The eventual formation of fractal-like patterns is due to particle clusters which may form due to shear-induced migration, lubrication forces between the colloids and capillary attraction at the oil-water interface.[58] Similar to what was observed by Li et al. during the withdrawal of a particle-liquid mixture from a hele-shaw cell, the radial retraction velocity gets reduced along the clusters leading to the formation of radial concentrated regions of deposited colloids.[58] With further retraction, the fractals disappear resulting in a compact colloid layer at the droplet base. However, after $t_E \approx 132$ min, we argue that the colloids cannot get further compacted and a 'friction-dominated regime' starts. At this regime, the frictional interaction of the accumulated colloids at the droplet base with the confining boundary (i.e. the droplet base-oil interface area) becomes so large that the boundary along with the colloidal deposit layer gets deformed (unlike suspension flow through a tube with fixed boundary). As a result, the colloid assembly starts to rise along the contours of the droplet-oil interface, resulting in the formation of the skirt-like deposition (as shown in the schematic of figure 7G3).

As the oil layer in the case of a LISS 1 surface is adequately thick, the contact line continues to retract and pull up the particle-containing oil layer until the complete coverage of the entire droplet with the colloidal particles after $t_E \approx 208$ min. As the entire droplet surface gets covered with particles, it further reduces the rate of evaporation. Once the droplet surface gets fully covered with colloids, the retraction of the TPCL stops and subsequent evaporation of the drop continues in the CCR mode without any sudden increase in $d_D$. Complete colloidal coverage of the inner wetting ridge-droplet interface leads to the reduction in interfacial tension. While this reduction in the interfacial tension engenders gradual drainage of the wetting ridge (already discussed before), no flattening of the droplet is observed over LISS 1 arguably due to the presence of a thicker deposited colloid layer at the droplet base resulting in higher friction within the colloid layer near the wetting ridge. As the remaining liquid in the droplet continues to evaporate, there is not enough liquid to support the shape of the stiffed colloid skin covering the droplet and consequently, compressive stress generated at the interface leads to the formation of wrinkles in the colloid skin layer, which can be clearly seen in figure 7F. It is important to point out that we did observe the accumulation of some colloidal particles at the droplet free surface even over oil-free flat and patterned sticky hydrophobic surfaces, as can be seen in frames A3 and B3 of figure 2. However, in those cases, the formation of particle assembly on these surfaces does not start from the TPCL and is attributed to the random convective motion of the particles within the liquid droplet, which migrates to the free surface and gets stuck, as can be seen in frames a3 and b3 of figure 2 respectively.

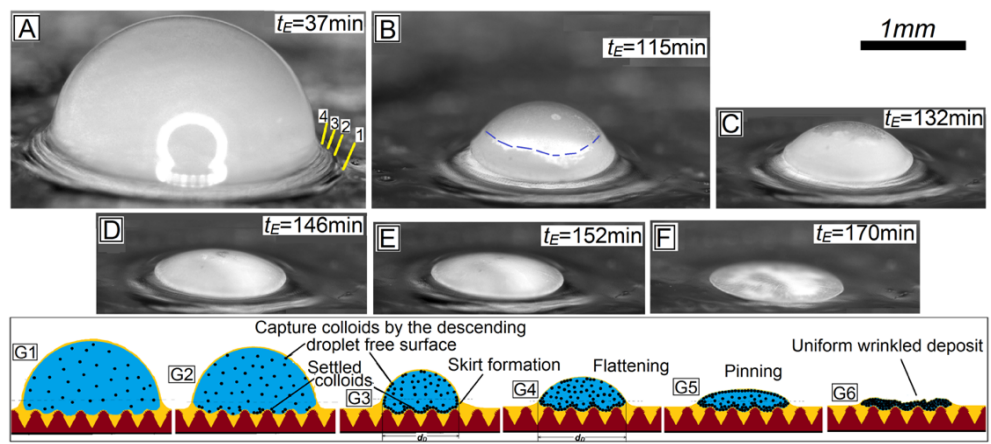

*Figure 8: Optical micrographs showing side-view images of an evaporating aqueous suspension droplet with $C_i \approx 0.075\%$ over a LISS 2 surface. Frames G1-G6 schematically show the evolution of colloidal self-assembly.*

Over a LISS 2 surface, the initial part of the evaporation and droplet evolution is similar to that observed over a LISS 1 surface, except for the formation of multiple overlapping wetting ridges as can be seen in figure 8A. Similar overlapping ridges were also observed during the evaporation of a pure water droplet over LISS 2 and LISS 3 surfaces, arguably due to a competition between the drainage of excess oil from the wetting ridge during contact line reorganization and slower retraction of the contact line due to the presence of the patterns of the submerged substrate. $h_D$ reduces with progressive evaporation leading to the drainage of oil from the encapsulating oil layer towards the base of the wetting ridge to balance the interfacial tension forces along the TPCL. And, after retraction of the TPCL, a new wetting ridge gets formed at the present location over the old wetting ridge which could not smoothen out due to restricted outward oil flow and forms multiple overlapping wetting ridges. The detailed reason for the formation of such multiple overlapping ridges can be found in our earlier publication.[37] However, the main difference between the evaporation of a drop of water and a colloidal suspension is the deposition of particles at the droplet interfaces. Similar to that observed over a LISS 1 surface, even over a LISS 2 surface and interfacial colloidal deposition starts in the form of skirt formation, which starts after $t_E \approx 115$ min, which is slightly earlier than that over a LISS 1 surface arguably due to lower cloaking layer thickness as a consequence of reduced $h_E$. Interestingly, no fractal-like colloidal assembly at the droplet base is observed in this case. We feel this is attributed to the undulating nature of the droplet base-oil water interface due to the presence of the micropillar at the base. Because the movement of the randomly deposited colloids at the droplet base occurs over an undulating oil-coated surface (as the droplet squeezes the oil in between the conical micropillars) in the initial 'viscous dissipation dominated regime' instead of an almost horizontal droplet base-oil interface during retraction of the periphery. However, once the droplet base gets covered with compact colloid assembly, the onset of skirt formation occurs due to the existence of a 'friction-dominated regime' similar to what was observed over LISS 1. However, unlike a complete growth observed over a LISS 1 surface, over a LISS 2 surface, the growth of the skirt/deposit region is limited only up to the height of the pre-existing wetting ridge as schematically explained in figures 8G3 and 8G4). This is because, after the formation of multiple wetting ridges, the contact line eventually gets pinned, due to localised thinning of the oil layer below the droplet, as oil rises along the droplet periphery towards the wetting ridge. As already discussed, due to a reduction in the interfacial tension, once the height of the deposition zone attains that of the wetting ridge, the wetting ridge itself starts to drain off after $t_E \approx 132$ min as can be seen in figure 8C. The drainage of the wetting ridge is associated with progressive flattening of the droplet with a slight increase in $d_D$ which continues till $t_E \approx 170$ min (figure 8F) by when the wetting ridge has completely drained off shifting the TPCL to its new pinned location. Interestingly, though the growth of the skirt-like deposition front on the droplet free surface stops as the TPCL gets pinned, the entire droplet eventually gets covered with colloidal particles, due to migration of the particles towards the free surface with progressive evaporations. This difference in the mechanism of how the droplet surface gets fully covered with colloids over different LISS surfaces is an extremely important observation of our study, which clearly highlights the role of the sticky base substrate in modulating the flow of oil along the contours of the droplet surface. Eventually, as observed over a LISS 1 surface, once the free surface of the droplet gets fully covered with particles at a later stage, with further evaporation of the liquid wrinkles are formed due to the generation of compressive stress in the stiffed colloid skin at the droplet free surface. However, due to early pinning during retraction, $d_F$ is higher over a LISS 2 surface than that observed over a LISS 1 surface.

The effect of varying $h_E$ on the final deposition patterns can be correlated with the rate of evaporation which in turn influences the dominant mechanism of colloidal self-assembly over each LISS surface. It important to highlight that the system involves two types of diffusion (1) diffusion of water vapor from the droplet free surface to the environment and (2) diffusion of colloids within the droplet due to evaporation mediated flow. Further due to the presence of the wetting ridge, as already pointed out the rate of evaporation is not maximum at the periphery leading to the absence of the radially outward advective flow. Assuming a diffusion-limited evaporation from the droplet free surface with no evaporation from the region near TPCL covered by the wetting ridge, Sharma et al. derived the following equation,[59]

$$R_b^2 \approx R_0^2 - \frac{4\lambda t_E \sin^2\theta_b f(\theta_{app})}{\pi \beta(\theta_b)} \quad \text{------------------ (1)}$$

Where $R_b$ and $R_0$ are the instantaneous droplet base radius and initial droplet base radius; $\theta_b$ and $\theta_{app}$ are the apparent contact angle at the droplet base (at the solid-TPCL) and apparent contact angle at the pseudo-liquid TPCL; $\beta(\theta_b) = (1 - \cos\theta_b)^2(2 + \cos\theta_b)$; and f($\theta_{app}$) is a polynomial function of $\theta_{app}$.[59]

To analyse the initial evaporation rate until the onset of pinning or flattening of the droplet, we have theoretically calculated the slope ($S_t$) of the $R_b^2$ vs $t_E$ plot using equation (1) for the evaporation of the suspension droplets of varying $C_i$ over different LISS surfaces as shown in the following table 2 and compared it with the experimental slope ($S_e$) calculated from the $R_b^2$ vs $t_E$ plots using the experimentally observed data.

*Table 2: Comparison of theoretical slope ($S_t$) and experimental slope ($S_e$) of the $R^2$ vs $t_E$ plot*

| Substrate | $C_i$ (%) | $\theta_b$ (°) | $\theta_{app}$ (°) | $\theta_{app}$ (rad) | $\beta(\theta_b)$ | $\sin^2\theta_b$ | $f(\theta_{app})$ | $-10^{10} * S_t$ (m²/s) | $-10^{10} * S_e$ (m²/s) |
|---|---|---|---|---|---|---|---|---|---|
| LISS 1 | 0 | 96.4 | 78.3 | 1.366 | 2.33 | 0.99 | 0.445 | 4.69 | 1.60 |
| | 0.075 | 95.9 | 73.2 | 1.277 | 2.31 | 0.99 | 0.420 | 4.46 | 2.03 |
| | 1.25 | 91.2 | 64.5 | 1.125 | 2.06 | 0.99 | 0.374 | 4.46 | 5.31 |
| LISS 2 | 0 | 108.4 | 96.6 | 1.685 | 2.91 | 0.90 | 0.527 | 4.04 | 3.18 |
| | 0.015 | 111.2 | 97.1 | 1.694 | 3.03 | 0.86 | 0.529 | 3.72 | 2.91 |
| | 0.075 | 110.1 | 95.9 | 1.673 | 2.81 | 0.88 | 0.524 | 4.07 | 2.89 |
| | 1.25 | 112.9 | 97.2 | 1.696 | 3.10 | 0.84 | 0.530 | 3.56 | 3.01 |
| LISS 3 | 0 | 113.8 | 102.2 | 1.783 | 3.14 | 0.84 | 0.549 | 3.64 | 3.84 |
| | 0.075 | 113.5 | 103.9 | 1.813 | 3.13 | 0.84 | 0.556 | 3.70 | 3.67 |
| LISS 4 | 0 | 113.4 | 106.3 | 1.855 | 3.12 | 0.84 | 0.564 | 3.77 | 3.98 |
| | 0.075 | 112.5 | 105.9 | 1.848 | 3.09 | 0.85 | 0.563 | 3.85 | 4.57 |
| | 1.25 | 113.5 | 96.7 | 1.687 | 3.13 | 0.84 | 0.528 | 3.96 | 5.77 |

The theoretically predicted slope ($S_t$) is comparable with the experimentally observed slope ($S_e$) for the evaporation of suspension droplets with $C_i \approx 0.075\%$ over LISS 3 and LISS 4, whereas over LISS 1 and LISS 2, $S_e$ is observed to be significantly smaller than $S_t$. This observation clearly highlights that the evaporation is diffusion-limited over LISS 3 and LISS 4 with only negligible influence of the encapsulating oil layer. This happens as an inherently thin encapsulating oil layer becomes even thinner due to the reduced interfacial tension at the droplet free surface by the presence of isolated colloidal islands which eventually evolves into a continuous colloid skin there. Additionally, over LISS 3 and LISS 4 surfaces, a higher rate of evaporation results in a higher value of the descend velocity ($V_i$) of the droplet-encapsulating oil layer interface (or the droplet free surface) ($V_i$ is calculated from the progressive variation of droplet height ($h_D$) with $t_E$) such that $0.5 < \frac{V_i}{V_p} < 2.0$ (Where $V_p = 2\sqrt{\frac{Dt}{\pi}}$ is the particle diffusion velocity; D is diffusion constant from Stokes-Einstein relation i.e. $D = \frac{k_B T}{6\pi\eta r}$) where according to Li et al.,[46] both $V_i$ and $V_p$ plays a significant role in the final deposit morphology. However, over a LISS surface, in the absence of the radially outward flow, although $V_p$ is comparable with $V_i$, the particles do not migrate towards the periphery and with progressive $t_E$ they settle at the droplet base. This corroborates with our experimental observation of both 'capture of colloids by the droplet free surface' and 'settling of colloids at the droplet base' mechanism contributing to the final deposition morphology over LISS 3 and LISS 4 surface. Interestingly, over a LISS 4 after almost complete coverage of the droplet free surface with colloids followed by pinning of the droplet, as the inherently thin wetting ridge almost disappears (after $t_E \approx 76$ min, figure 3C5), a higher evaporation rate near periphery leads to a thin peripheral ring formation.

Over LISS 1 and LISS 2 surfaces, a thick encapsulating oil layer exists over the droplet free surface resulting in reduced evaporation rate which leads to a significantly smaller $S_e$ compared to $S_t$ (table 2), as equation 1 does not include the effect of an encapsulating oil layer. Additionally, over LISS 2, multiple overlapping wetting ridges leads to a significant increase in instantaneous wetting ridge height ($h_R$) compared to the equilibrium value of $h_R$ corresponding to a specific droplet height ($h_D$) during evaporation. This results in a smaller area above the wetting ridge which is available for evaporation thereby reducing the evaporation rate and $S_e$. Over LISS 1 and LISS 2 surfaces, a slower evaporation rate leads to a smaller interfacial velocity ($V_i$) of the descending droplet free surface such that $\frac{V_i}{V_p}$ is < 0.5 (table 2) and according to Burkhert et al.,[62] $V_p$ dominates. This matches our experimental observation over LISS 1 and LISS 2 surfaces with only a few colloids getting captured by the descending droplet free surface whereas most the colloids eventually deposit at the droplet base followed by the transfer of them from the droplet base to the droplet apex via skirt-like deposition. It is important to note that we could not capture the effect of colloidal assembly at the droplet-wetting ridge interface on the evaporation rate for different LISS surfaces as it leads to the drainage of the wetting ridges resulting in the increase in $R_b$ which reverses the slope of the $R_b^2$ vs $t_E$ plot over all LISS surfaces except LISS 1 where $R_b$ remains constant.

### 3.2 Effect of $C_i$ on evaporation dynamics and deposit patterns

Once the effect of $h_E$ on the evaporation dynamics and final deposition morphology is well understood for suspension droplets with $C_i \approx 0.075\%$ over different LISS surfaces, we now explore the effect of initial colloid concentration ($C_i$) in detail. Experiments were performed for $C_i \approx 0.015\%$ and $C_i \approx 1.25\%$ over LISS 2 surface where the effect of slipperiness as well as that of the underlying substrate patterns on the dynamics of the droplet is simultaneously manifested. Different frames of figure 9 show the evolution of the evaporating droplet of $C_i \approx$ 0.015% (figure 9 A) and $C_i \approx 1.25\%$ (figure 9 B) respectively, as a function of $t_E$ over LISS 2 surfaces. The graphs in frames A, B and C of figure 10 show how $d_D$, $h_D$ and $\theta_{app}$ of the droplet progressively change with $t_E$. To facilitate the comparison between different $C_i$, we have also plotted the data for $C_i \approx 0.075\%$ in all frames of figure 8 as well as that of pure water (i.e. $C_i \approx$ 0%). The initial apparent contact angle ($\theta_{app,i} \approx 96.6° \pm 1.0°$) and initial wetting ridge dimensions are almost independent of $C_i$, as can be seen in figure 10C.

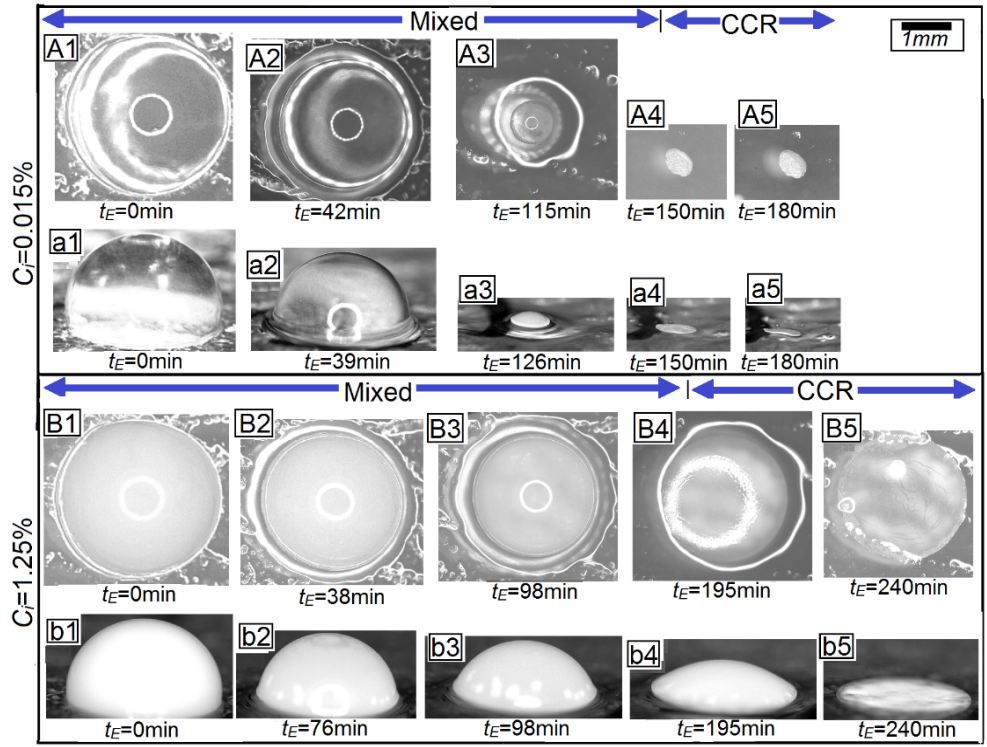

*Figure 9: Optical micrographs (top and side view) showing the temporal evolution of evaporating aqueous suspension droplets of different $C_i$ over a LISS 2 surface: (A, a) $C_i \approx 0.015\%$ and (B, b) $C_i \approx 1.25\%$ (Scale bar: 1mm).*

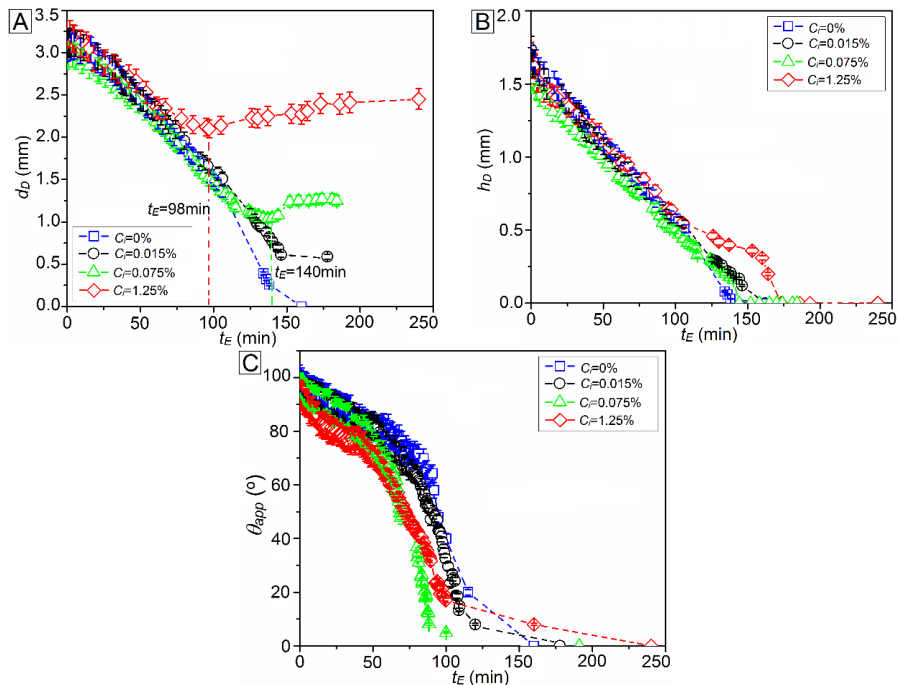

*Figure 10: Plot of droplet (A) diameter ($d_D$), (B) height ($h_D$) and (C) apparent contact angle ($\theta_{app}$) of evaporating aqueous suspension droplets with varying $C_i$ as a function of $t_E$ over a LISS 2 surface.*

For a dilute suspension droplet with $C_i \approx 0.015\%$, the droplet retracts with progressive evaporation and reduction in droplet volume with hardly any colloids being captured at the droplet free surface over almost 70% of the evaporation duration. Also, the system is dominated by $V_p$ as $\frac{V_i}{V_p} \approx 0.42$ (table 2) which implies that the most of the colloids get deposited at the droplet base rather than getting captured by the descending droplet free surface. After $t_E \approx 120$ min, the deposited colloidal assembly at the droplet base starts to climb along the inner contour of the wetting ridge, forming a skirt-like region which has already been discussed for $C_i \approx 0.075\%$. Beyond this stage however, the droplet free surface rather quickly gets fully covered with colloids within $t_E \approx 126$ min (frame a3, figure 9) and the wetting ridge starts to drain off due to a reduction in interfacial tension resulting in the pinning of the droplet solid-TPCL.

On the other hand, for $C_i \approx 1.25\%$, a large number of colloids get captured fast at the droplet free surface forming isolated islands due to the availability of a much higher number of colloids For this system, as $\frac{V_i}{V_p} \approx 0.52$ (table 2), the deposition mechanism is attributed to both settling of particles at the droplet base and capture of the particles by the droplet free surface. The size of the colloidal islands grows with time eventually covering it completely. While similar coverage of the droplet free surface was earlier observed in a droplet with $C_i \approx 0.075\%$ over LISS 1 surface, the precise mechanism of the interfacial colloidal coverage is completely different in the two cases. In the previous case, it was due to the growth of the skirt region. In contrast, in the present case, the surface coverage is due to the direct capture of the colloids by the descending droplet free surface (or the droplet-encapsulating layer interface). This results in a complete surface coverage with particles at a higher droplet radius. Beyond $t_E \approx 98$ min, a slower rate of decrease of $h_D$ is observed up to $t_E \approx 150$ min in a drop with $C_i \approx 1.25\%$ which can be seen in figure 10B due to the reduced rate of evaporation as the free surface gets fully covered with particles. This reduces the retraction velocity of the contact line. Thus with the increase in $C_i$, $t_F$ increases and a larger final deposit is formed as can be seen from the last column of figure 9 and figure 2 C5. An almost uniform colloidal deposit without any peripheral ring along with wrinkles at the free surface is observed during the evaporation of colloidal suspension droplets over a LISS 2 surface for $C_i \approx 0.075\%$, as can be seen from figure 9 and figure 2 C. However, for dilute suspension droplets of $C_i \approx 0.015\%$, wrinkles eventually disappear after complete evaporation as can be seen in figure 9 A5 as they collapse over the top of the micropillars of the base substrate and rearrange themselves by capillary interaction. In contrast, permanent wrinkles are observed in the final deposit pattern obtained for $C_i \approx 1.25\%$ (figure 9 B5) as in this case the base patterned substrate gets buried under a thick colloid layer deposited at the droplet base during evaporation. So, the wrinkle morphology remains unphased by the micropillars of the patterned base substrate. Further, a significantly smaller $S_e$ compared to $S_t$ is obtained for $C_i \approx 0.015\%$ because of the multiple overlapping wetting ridges (table 2). On the other hand, for $C_i \approx 1.25\%$, due to availability of large number of colloids, a rapid rate of capture of colloids by the droplet free surface is observed resulting in a significant reduction in interfacial tension at the droplet free surface. This leads to the drainage of the encapsulating oil layer and a significantly thinner encapsulating oil layer enhancing the evaporation rate which balances the reducing effect of multiple overlapping wetting ridges on the evaporation rate resulting in a $S_e$ value comparable with $S_t$ (table 2).

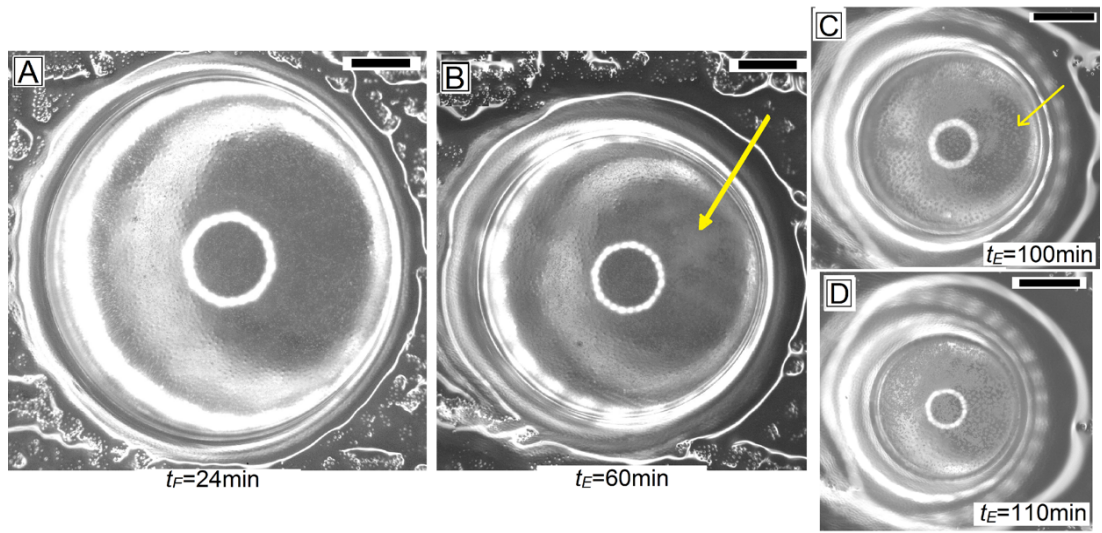

*Figure 11: Optical micrographs of colloid deposition and self-assembly at the droplet base–oil interface during evaporation of an aqueous suspension droplet with $C_i \approx 0.015\%$ over a LISS 2 surface (Scale bar: 500μm).*

Next, we focus on the nature of the colloidal self-assembly at the droplet base with progressive evaporation and retraction of the TPCL for $C_i \approx 0.015\%$. In fact, in this particular case as $C_i$ is low the crowding and coverage of the droplet free surface with colloids are slow, which in turn allowed us to directly view the droplet base. Such a study was not possible at higher $C_i$, as the colloidal accumulation at the droplet free surface obstructed the view of the optical microscope. We observe that the colloids initially deposit randomly at the droplet base–oil interface and eventually form large colloidal patches driven by capillary attraction[49] after $t_E \approx 60$ min (figure 11B). With subsequent evaporation and retraction of the solid-TPCL, some of the previously deposited colloids get attached to the large colloidal patch while some other particles rearrange themselves near over the tops of the conical micropillars after $t_E \approx 100$ min (figure 11C) due to the well-known 'Cheerios effect'.[49-50] There are two forces i.e., buoyancy force and the surface tension force acting on the colloids settled at the droplet base-oil interface. As the density of oil ($\approx 970$ kg/m$^3$) is less than that of the PS colloids ($\approx 1050$ kg/m$^3$)[48], the buoyancy force tends to push the colloids into the oil layer whereas the higher surface tension of water tends to keep the colloid within the water layer. This results in the assembly of the PS colloids partially immersed into the oil layer deforming the oil–water interface at the droplet base. Further, as the droplet penetrates in between the micropillars by squeezing out the oil and forming a 'slippery Wenzel' wetting state, there is a curvature of the oil-water interface near each micropillar top. Therefore, an effective upward force along with the capillary interactions acts on the settled colloids to drive them upward along the curvature of the oil-water interface towards a neighbouring micropillar top. The number of colloids assembled near each micropillar increases with $t_E$ up to $t_E \approx 110$ min (figure 11 D) beyond which these small colloidal patches over each micropillar eventually get close to each other and merge to become one large assembly. A similar type of colloid self-assembly is also observed for other values of $C_i$ over a LISS 2 surface and for $C_i \approx 0.075\%$ over all the LISS surfaces except LISS 1.

Interestingly, although after complete evaporation the final colloidal deposit sinks into the oil layer irrespective of $C_i$ and $h_E$, a thin layer of oil remains in between the colloidal assembly and the substrate patterns underneath the oil. This happens as the spreading coefficient ($S_{REP}$) for the replacement process of the oil layer by the PS colloids is negative.[51]

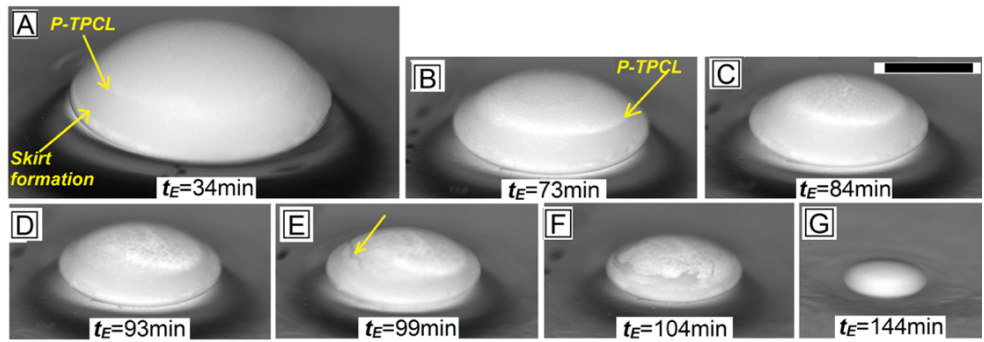

*Figure 12: Optical micrographs showing side view images of an evaporating aqueous suspension droplet with $C_i \approx 1.25\%$ over a LISS 1 surface between $t_E \approx 34$ min and $t_E \approx 144$ min. Pseudo-liquid TPCL (PTPCL) and skirt region are marked in frames A and B. Crack in the colloid skin at the droplet free surface is marked in frame E (Scale: 1 mm).*

After understanding the role of $C_i$ on the deposition morphology over LISS2 surfaces, we now investigate the same over oil-flooded LISS 1 and oil-starved LISS 4 surfaces. Figure 12 shows the evolution of colloidal self-assembly at different interfaces during the evaporation of an aqueous suspension droplet of $C_i \approx 1.25\%$ over a LISS 1 surface. The onset of 'skirt formation' happens significantly faster over a LISS 1 at higher $C_i$ (figure 12A), as a consequence of higher availability of oil as well as availability of a larger number of colloids. This in turn results in a faster formation of a compact colloid layer at the droplet base with the settling of more colloids along with a faster retraction velocity of the contact line as $\frac{V_i}{V_p} \approx 0.40$ which is < 0.5 (table 2).

Consequently, the final deposition is dominated by the settling of colloids at the droplet base. The higher retraction velocity of the contact line is a consequence of a significantly large number of colloids getting arrested by the descending droplet free surface resulting in a thinner encapsulating oil layer and a comparatively higher evaporation rate compared to what was observed in figures 3 and 4 for dilute suspension droplets over a LISS 1 (table 2). With progressive evaporation and retraction, eventually, the droplet free surface gets fully covered with colloids which is followed by wrinkling of the colloid skin above the pseudo-liquid TPCL (P-TPCL) after $t_E \approx 84$ min (figures 12C and 12D). In the meantime, the wetting ridge-droplet interface also gets covered with colloids and the wetting ridge starts to drain out. At this stage, the wrinkled colloidal assembly at the droplet free surface cannot withstand the compressive stress any further due to the reduction in drop volume and cracks get formed on it (figure 12E). Eventually, with further evaporation and retraction, the skirt region and the colloidal islands at the droplet free surface fully cover the droplet and form a pallet-like final deposit (figure 12G). In addition to a thinner encapsulating layer due to colloidal assembly at the droplet free surface, the drainage of the wetting ridge induced by the colloidal assembly at the wetting ridge-droplet interface leads to a slightly higher $S_e$ than $S_t$ for $C_i \approx 1.25\%$ unlike what is observed for dilute suspension droplets over LISS 1 (table 2).

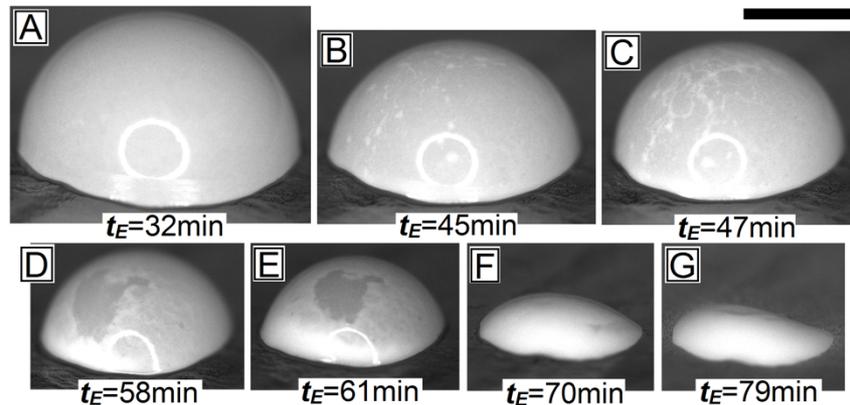

Figure 13: Optical micrographs showing side view images of an evaporating aqueous suspension droplet with $C_i \approx 1.25\%$ over a LISS 4 surface between $t_E \approx 32$ min and $t_E \approx 79$ min (Scale: 1 mm).

Interestingly, in the case of evaporation of concentrated suspension droplet with $C_i \approx 1.25\%$ over LISS 4 surface, an almost uniform colloidal deposit with wrinkles at the droplet free surface as can be seen in figure 10 is observed instead of a final deposit with a thin peripheral ring seen for $C_i \approx 0.075\%$ (figure 3C). In the case, as $\frac{V_i}{V_p} \approx 0.75$ (> 0.5; table 2), colloids get captured by the descending droplet free surface along with deposition at the droplet base. However, due to a higher value of $C_i$, the skin formation is expedited, as it forms completely after $t_E \approx 32$ min (compared to a slower skin formation after $t_E \approx 64$ min for 0.075% over LISS 4 in figure 3C). With progressive evaporation and retraction, the colloid skin gets wrinkled and eventually gets torn locally near the contact line (frames B-D of figure 13). However, in the meantime, a large number of the particles also get deposited all over the droplet base, leading to an almost uniform final deposit morphology. Also, in this case, $S_e$ is significantly higher than $S_t$ for $C_i \approx 1.25\%$ over LISS 4 (table 2) arguably due to an almost negligible encapsulating oil layer (due to a faster formation of colloid skin at the droplet free surface) and evaporation in a series of alternative mixed mode and CCR mode evaporations due to a stick-slip type retraction (as can be seen in frames A-D of figure 13). According to McHale et al., the equation (1) is applicable for mixed-mode or CCA mode.[60] The intermediate CCR modes due to temporary pinning of the periphery occur as a combined effect of a large number of settled colloids at the droplet base and an undulating surface at the droplet base of the oil-starved LISS 4 surface. Because of this, there is also no flattening of the droplet is observed. Further, for $C_i \approx 1.25\%$, $S_e$ is significantly higher than what is observed for $C_i \leq 0.075\%$ over all the LISS surfaces (table 2) due to the faster formation of colloidal islands at the droplet-wetting ridge interface and at the droplet free surface resulting in a smaller wetting ridge and thinner encapsulating oil layer for most of the evaporation duration.

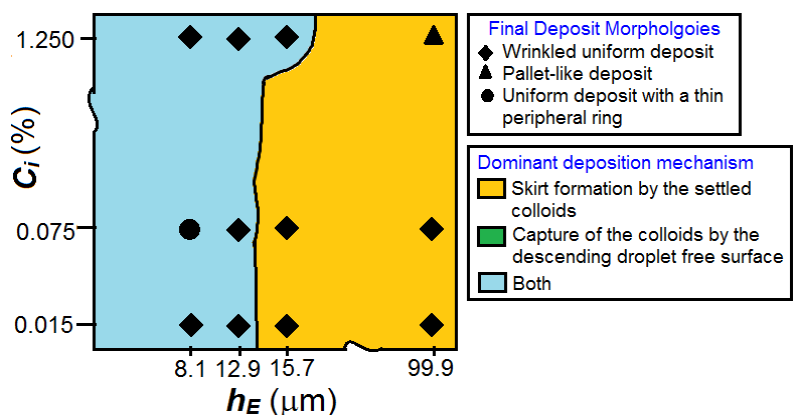

*Figure 14: A phase diagram showing the effect of infusing oil layer thickness ($h_E$) and initial colloid concentration ($C_i$) on the final colloidal deposition morphologies and the dominant mechanism for the formation of final deposit.*

The overall deposition patterns for suspension droplets with varying $C_i$ over different LISS surfaces are presented in a phase diagram in figure 14. An almost uniform final colloidal deposit with wrinkles and with no peripheral ring is observed in most of the cases for suspension droplets with varying $C_i$ on different LISS surfaces except a uniform deposit with a peripheral ring for $C_i \approx 0.075\%$ over LISS 4 and a pallet-like deposit for $C_i \approx 1.25\%$ over LISS 1. Also, as discussed earlier, depending on the value of $C_i$ and $h_E$, there are two different mechanisms leading to the final colloidal deposit which are skirt formation by the settled colloids and the capture of colloids by the descending droplet free surface as shown in the phase diagram of figure 14. Over LISS 1 surface, most of the colloids get settled at the droplet base followed by the formation of a skirt-like region irrespective of the value of $C_i$ (figures 7 and 12 for $C_i \approx 0.075\%$ and 1.25% respectively). Although for $C_i \approx 1.25\%$ over LISS 1, a colloid skin gets formed by the captured colloids by the droplet free surface with smaller area at a later stage as shown in figure 12 C, most of the colloids deposit at the droplet base and onset of skirt formation occurs fast only after $t_E \approx 34$ min and continues to grow with $t_E$. A similar deposition dynamics is also observed for $C_i \approx 0.075\%$ (figure 2C) and 0.015% (figure 9A) over LISS 2 and LISS 3 surfaces as can be seen in figure 14. Interestingly, for all values of $C_i$ over the LISS 3 and 4 surface, the final deposit pattern is attributed to both mechanisms as can be seen in figure 14 due to a rapid evaporation rate leading to a higher value of the interfacial velocity of the droplet free surface as the encapsulating oil layer and the wetting ridge are arguably very thin.

4. **Conclusions**

We have presented a detailed experimental account of the evaporative drying of a colloidal droplet over a slippery surface. The presence of the sticky surface patterns below the slip-enhancing oil layer ensures the stability of the oil layer for all $h_E$ and significantly modulates the availability and dynamics of oil in the wetting ridge and the cloaking layer. Depending on the type of LISS surfaces used (with different $h_E$) and $C_i$ of the suspension droplets, the deposition is either dominated by the settling of the colloids at the droplet base followed by skirt formation (when $V_i/V_p < 0.5$)[46,62] or the capture of colloids by the descending droplet free surface (When $V_i/V_p > 2.0$)[46,62] or both (transition regime when $0.5 < V_i/V_p < 2.0$)[46,62] which depends on the ratio of $V_i/V_p$. The unique features observed during progressive evaporation include the formation of a skirt at the TPCL rising along the wetting ridge-oil interface where the rise of the skirt is favoured over LISS surfaces with higher $h_E$. The second major observation is the coverage of the evaporating droplet free surface by colloidal skin. Here again, our experiments reveal that the mechanism of skin formation may be drastically different depending on a combination of $h_E$ and $C_i$. For example, we have shown that for $C_i \approx 0.075\%$ over a LISS 1 surface, the skin formation is entirely due to the growth of the skirt. In contrast, the skin formation over a LISS 4 for $C_i \approx 0.075\%$ is primarily due to the capture of the colloids by the descending droplet free surface. In contrast, we have also captured situations where both these collectively led to the formation of colloid skin (for $C_i \approx 0.075\%$ over LISS 2). Here we also show that although the final deposit pattern in all cases is attributed to the collapse of the skin layer with $t_E$ after complete evaporation of water, the precise morphology depends on $h_E$ and $C_i$ which has been shown in the phase diagram (figure 14). Thus we have presented here an extremely novel evolution pathway during the evaporation of colloidal suspension droplets over slippery surfaces. While the work is predominantly fundamental, the findings reported in this article can be used to understand the efficiency of different commercial slippery surfaces fabricated for outdoor purposes. This work also highlights how an almost uniform colloidal assembly can be obtained over liquid-infused sticky surfaces (LISS) which on a bare sticky surface is otherwise not possible. An array of microliter suspension droplets on such LISS surfaces can be utilised as a bioassay for biomolecular detection,[52] antimicrobial susceptibility testing and so on.[53]

**NOMENCLATURE:**

**Abbreviations:**

LISS: Liquid-infused slippery substrate

LIS: Lubricant-impregnated surfaces

SLIPS: Slippery liquid-infused porous surfaces

TPCL: Three-phase contact line

CAH: Contact angle hysteresis

CCA: Constant-contact angle

CCR: Constant contact radius

**Symbols:**

$\theta_{app}$: Apparent Contact Angle

$\theta_{app, i}$: Initial apparent Contact Angle

$C_i$: Initial droplet concentration

$h_E$: Effective oil layer thickness

$h_p$: Height of conical micropillars

$\gamma_{sa}$: Solid-air interfacial tension

$\gamma_{wa}$: ater-air interfacial tension

$\gamma_{os}$: Oil-solid interfacial tension

$\gamma_{oa}$: Oil-air interfacial tension

$\gamma_{sw}$: Solid-water interfacial tension

$\gamma_{ow}$: Oil-water interfacial tension

$\theta_{os(a)}$: Contact angle of an oil drop on a flat solid substrate in air

$S_{ow(a)}$: Spreading coefficient of oil on water in the air

$A_H$: Hamaker constant

$\gamma_{LV}$: liquid-vapour interfacial tension

$d_D$: Apparent droplet diameter

$h_D$: Droplet height

$R_b$: Droplet base radius

$R_0$: Initial droplet base radius

$d_F$: Final deposit diameter

$t_F$: Total lifetime of an aqueous colloid suspension drop

$t_{F0}$: Total lifetime of a pure water drop

$h_R$: Height of the wetting ridge

$w_R$: Width of the wetting ridge

**Reference**

1. B. -J. de Gans, U. S. Schubert, Inkjet Printing of Well-Defined Polymer Dots and Arrays, Langmuir 20 (2004) 7789-7793.

2. D. -D. Li, L. Wang, J. Liu, Z. Huang, Manipulating Nano-suspension Droplet Evaporation by Particle Surface Modification, Langmuir 37 (2021) 12234−12241.
3. J. R. Trantum, M. L. Baglia, Z. E. Eagleton, R. L. Mernaugh, F. R. Haselton, Biosensor design based on Marangoni flow in an evaporating drop, Lab Chip 14 (2014) 315-324.
4. F. R. Smith, D. Brutin, Wetting and spreading of human blood: Recent advances and applications, Curr Opin Colloid Interface Sci 36 (2018) 78–83.
5. F. R. Smith, C. Nicloux, D. Brutin, A new forensic tool to date human blood pools, Sci. Rep. 10 (2020) 8598 (1-12).
6. M. Usman, X. Guo, Q. Wu, J. Barman, S. Su, B. Huang, T. Biao, Z. Zhang, Q. Zhan, Facile silicone oil-coated hydrophobic surface for surface enhanced Raman spectroscopy of antibiotics, RSC Adv. 9 (2019) 14109-14115.
7. R. G. Picknett, R. Bexon, The evaporation of sessile or pendant drops in still air, J. Colloid and Interface Sci. 61 (1976) 336 – 350.
8. H. Hu, R. G. Larson, Marangoni Effect Reverses Coffee-Ring Depositions, J. Phys. Chem. B 110 (2006) 7090–7094.
9. R. D. Deegan, O. Bakajin, T. F. Dupont, G. Huber, S. R. Nagel, T. A. Witten, Contact line deposits in an evaporating drop, Phys. Rev. E 62 (2000) 756-765.
10. R. D. Deegan, O. Bakajin, T. F. Dupont, G. Huber, S. R. Nagel, T. A. Witten, Capillary flow as the cause of ring stains from dried liquid drops, Nature 389 (1997) 827–829.
11. D. Orejon, K. Sefiane, M. E. R. Shanahan, Stick Slip of Evaporating Droplets: Substrate Hydrophobicity and Nanoparticle Concentration, Langmuir 27 (2011) 12834–12843.
12. L. Bansal, P. Seth, S. Sahoo, R. Mukherjee, S. Basu, Beyond coffee ring: Anomalous self-assembly in evaporating nanofluid droplet on a sticky biomimetic substrate, Appl. Phys. Lett. 113 (2018) 213701(1-5).
13. Y.-F. Li, Y.-J. Sheng, H.-K. Tsao, Evaporation Stains: Suppressing the Coffee-Ring Effect by Contact Angle Hysteresis, Langmuir 29 (2013) 7802−7811.
14. N. D. Patil, P. G. Bange, R. Bhardwaj, A. Sharma, Effects of Substrate Heating and Wettability on Evaporation Dynamics and Deposition Patterns for a Sessile Water Droplet Containing Colloidal Particles, Langmuir 32 (2016) 11958−11972.
15. A. Gao, J. Liu, L. Ye, C. Schönecker, M. Kappl, H. J. Butt, W. Steffen, Control of Droplet Evaporation on Oil-Coated Surfaces for the Synthesis of Asymmetric Supraparticles, Langmuir 35 (2019) 14042−14048.
16. J. Kim, H. Hwang, H. J. Butt, S. Wooh, Designing the shape of supraparticles by controlling the apparent contact angle and contact line friction of droplets, J. Colloid and Interface Sci. 588 (2021) 157–163.
17. Y. Li, C. Diddens, T. Segers, H. Wijshoff, M. Versluis, D. Lohse, Evaporating droplets on oil-wetted surfaces: Suppression of the coffee-stain effect, Proc. Natl. Acad. Sci. U. S. A. 117 (2020) 16756-16763.
18. J. D. Smith, R. Dhiman, S. Anand, E. R. Garduno, R. E. Cohen, G. H. McKinley, K. K. Varanasi, Droplet mobility on lubricant-impregnated surfaces, Soft Matter 9 (2013) 1772-1780.
19. N. Bjelobrk, H. L. Girard, S. B. Subramanyam, H. M. Kwon, D. Quere, K. K. Varanasi, Thermocapillary motion on lubricant-impregnated surfaces, Phys Rev Fluids 1 (2016) 063902(1-7).
20. X. Dai, B. B. Stogin, S. Yang, T. S. Wong, Slippery Wenzel State, ACS Nano 9 (2015) 9260-9267.
21. S. A. McBride, S. Dash, K. K. Varanasi, Evaporative Crystallization in Drops on Superhydrophobic and Liquid-Impregnated Surfaces, Langmuir 34 (2018) 12350−12358.
22. S. Bandyopadhyay, S. M. Sriram, V. Parihar, S. Das Gupta, R. Mukherjee, S. Chakraborty, Tunable adhesion and slip on a bio-mimetic sticky soft surface, Soft Matter 15 (2019) 9031-9040.
23. D. Zhang, Y. Xia, X. Chen, S. Shi, L. Lei, PDMS-Infused Poly(High Internal Phase Emulsion) Templates for the Construction of Slippery Liquid-Infused Porous Surfaces with Self-cleaning and Self-repairing Properties, Langmuir 35 (2019) 8276 − 8284.

24. T.-S. Wong, S. H. Kang, S. K. Y. Tang, E. J. Smythe, B. D. Hatton, A. Grinthal, J. Aizenberg, Bioinspired self-repairing slippery surfaces with pressure-stable omniphobicity, Nature 477 (2011) 443 – 447.
25. L. Xiao, J. Li, S. Mieszkin, A. D. Fino, A. S. Clare, M. E. Callow, J. A. Callow, M. Grunze, A. Rosenhahn, P. A. Levkin, Slippery Liquid-Infused Porous Surfaces Showing Marine Antibiofouling Properties, ACS Appl. Mater. Interfaces 5 (2013) 10074–10080.
26. P. W. Wilson, W. Lu, H. Xu, P. Kim, M. J. Kreder, J. Alvarenga, J. Aizenberg, Inhibition of ice nucleation by slippery liquid-infused porous surfaces (SLIPS), Phys. Chem. Chem. Phys. 15 (2013) 581 – 585.
27. J. H. Guan, G. G. Wells, B. Xu, G. McHale, D. Wood, J. Martin, S. Stuart-Cole, Evaporation of Sessile Droplets on Slippery Liquid-Infused Porous Surfaces (SLIPS), Langmuir 31 (2015) 11781 − 11789.
28. F. Schellenberger, J. Xie, N. Encinas, A. Hardy, M. Klapper, P. Papadopoulos, H. J. Butt, D. Vollmer, Direct observation of drops on slippery lubricant-infused surfaces, Soft Matter 11 (2015) 7617 – 7626.
29. A. K. Epstein, T.-S. Wong, R. A. Belisle, E. M. Boggs, J. Aizenberg, Liquid-infused structured surfaces with exceptional anti-biofouling performance, Proc. Natl. Acad. Sci. U. S. A. 109 (2012) 13182–13187.
30. S. Bandyopadhyay, S. Santra, S. S. Das, R. Mukherjee, S. Chakraborty, Non-wetting Liquid-Infused Slippery Paper, Langmuir 37 (2021) 13627−13636.
31. V. Rastogi, O. D. Velev, Development and evaluation of realistic microbioassays in freely suspended droplets on a chip, Biomicrofluidics 1 (2007) 014107(1-17).
32. Y. Yu, H. Zhu, J. Frantz, M. Reding, K. Chan, H. Ozkan, Evaporation and coverage area of pesticide droplets on hairy and waxy leaves, Biosyst. Eng. 104 (2009) 324–334.
33. S. T. Chang, O. D. Velev, Evaporation-Induced Particle Microseparations inside Droplets Floating on a Chip, Langmuir 22 (2006) 1459–1468.
34. G. Chaniel, M. Frenkel, V. Multanen, E. Bormashenko, L. Chaniel, M. Frenkel, Paradoxical coffee-stain effect driven by the Marangoni flow observed on oil-infused surfaces, Colloids Surf. A: Physicochem. Eng. Asp. 522 (2017) 355-360.
35. M. Sharma, P. K. Roy, R. Pant, K. Khare, Sink dynamics of aqueous drops on lubricating fluid coated hydrophilic surfaces, Colloids Surf. A Physicochem. Eng. Asp. 562 (2019) 377–382.
36. B. Bhatt, S. Gupta, M. Sharma, K. Khare, Dewetting of non-polar thin lubricating films underneath polar liquid drops on slippery surfaces, J. Colloid and Interface Sci. 607 (2022) 530 – 537.
37. S. Sahoo, R. Mukherjee, Evaporative Drying of a Water droplet on Liquid Infused Sticky Surfaces, Colloids Surf. A Physicochem. Eng. Asp. 657 (2023) 130514.
38. U. U. Ghosh, S. Nair, A. Das, R. Mukherjee, S. DasGupta, Replicating and resolving wetting and adhesion characteristics of a Rose petal, Colloids Surf. A: Physicochem. Eng. Asp. 561 (2019) 9–17.
39. D. Roy, K. Pandey, M. Banik, R. Mukherjee, S. Basu, Dynamics of droplet impingement on bioinspired surface: insights into spreading, anomalous stickiness and break-up, Proc. R. Soc. A 475 (2019) 20190260 (1-22).
40. S. Roy, R. Mukherjee, Ordered to Isotropic Morphology Transition in Pattern-Directed Dewetting of Polymer Thin Films on Substrates with Different Feature Heights, ACS Appl. Mater. Interfaces 4 (2012) 5375–5385.
41. N. Bhandaru, R. Mukherjee, Ordering in Dewetting of a Thin Polymer Bilayer with a Topographically Patterned Interface, Macromolecules 54 (2021) 4517−4530.
42. S. -M. Park, M. P. Stoykovich, R. Ruiz, Y. Zhang, C. T. Black, P. F. Nealey, Directed Assembly of Lamellae-Forming Block Copolymers by Using Chemically and Topographically Patterned Substrates, Adv. Mater. 19 (2007) 607–611.
43. D. U. Ahn, Z. Wang, R. Yang, Y. Ding, Hierarchical polymer patterns driven by capillary instabilities at mobile and corrugated polymer–polymer interfaces, Soft Matter 6 (2010) 4900–4907.
44. Z. Zhang, D. U. Ahn, Y. Ding, Instabilities of PS/PMMA Bilayer Patterns with a Corrugated Surface and Interface, Macromolecules 45 (2012) 1972−1981.

45. S. Gerdes, G. Ström, Spreading of oil droplets on silicon oxide surfaces with parallel V-shaped channels, Colloids Surf. A: Physicochem. Eng. Asp. 116 (1996) 135-144.
46. Y. Li, Q. Yang, M. Li, Y. Song, Rate-dependent interface capture beyond the coffee-ring effect, Sci Rep 6 (2016) 24628 (1-7).
47. M. J. Kreder, D. Daniel, A. Tetreault, Z. Cao, B. Lemaire, J. V. I. Timonen, J. Aizenberg, Film Dynamics and Lubricant Depletion by Droplets Moving on Lubricated Surfaces, Phys Rev. X 8 (2018) 031053 (1-16).
48. W. Li, W. Ji, D. Lan, Y. Wang, Self-Assembly of Ordered Microparticle Monolayers from Drying a Droplet on a Liquid Substrate, J. Phys. Chem. Lett. 10 (2019) 6184−6188.
49. R. Mcgorty, J. Fung, D. Kaz, V. N. Manoharan, Colloidal self-assembly at an interface, Mater. Today 13 (2010) 34 – 42.
50. D. Vella, L. Mahadevan, The ''Cheerios effect'', Am. J. Phys. 73 (2005) 817 – 825.
51. A. Das, R. Mukherjee, Feature Size Modulation in Dewetting of Nanoparticle-Containing Ultrathin Polymer Films, Macromolecules 54 (2021) 2242−2255.
52. A. Shakeri, N. A. Jarad, J. Terryberry, S. Khan, A. Leung, S. Chen, T. F. Didar, Antibody Micropatterned Lubricant-Infused Biosensors Enable Sub-Picogram Immunofluorescence Detection of Interleukin 6 in Human Whole Plasma, Small 16 (2020) 2003844 (1-10).
53. X. Liu, X. Li, N. Wu, Y. Luo, J. Zhang, Z. Yu, F. Shen, Formation and Parallel Manipulation of Gradient Droplets on a Self-Partitioning SlipChip for Phenotypic Antimicrobial Susceptibility Testing, ACS Sens. 7 (2022) 1977−1984.
54. S. Roy, K. J. Ansari, S. S. K. Jampa, P. Vutukuri, R. Mukherjee, Influence of Substrate Wettability on the Morphology of Thin Polymer Films Spin-Coated on Topographically Patterned Substrates. ACS Appl. Mater. Interfaces 4 (2012) 1887−1896.
55. H. F. Bohn, W. Federle, Insect aquaplaning:Nepenthespitcher plants captureprey with the peristome, a fully wettablewater-lubricated anisotropic surface. Proc. Natl. Acad. Sci. U. S. A. 101 (2004) 14138–14143.
56. G. Dumazer, B. Sandnes, M. Ayaz, K. J. Måløy, E. G. Flekkøy, Frictional Fluid Dynamics and Plug Formation in Multiphase Millifluidic Flow. PRL 117 (2016) 028002 (1-5).
57. G. Dumazer, B. Sandnes, K. J. Måløy, E. G. Flekkøy, Capillary bulldozing of sedimented granular material confined in a millifluidic tube. Phys. Rev. Fluids 5 (2020) 034309 (1-18).
58. D. Li, Z. Yang, R. Zhang, R. Hu, Y. -F. Chen, Morphological patterns and interface instability during withdrawal of liquid-particle mixtures. J. Colloid Interface Sci. 608 (2022) 1598–1607.
59. M. Sharma, S. S. Mondal, P. K. Roy, K. Khare, Evaporation dynamics of pure and binary mixture drops on dry and lubricant coated slippery surfaces. J. Colloid Interface Sci. 569 (2020) 244 – 253.
60. G. McHale, S. M. Rowan, M. I. Newton, M. K. Banerjee, Evaporation and the Wetting of a Low-Energy Solid Surface. J. Phys. Chem. B 102 (1998), 1964-1967.
61. D. Xiong, Y. Qin, J. Li, Y. Wan, R. Tyagi, Tribological properties of PTFE/laser surface textured stainless steel under starved oil lubrication. Tribology International 82 (2015) 305–310.
62. C. T. Burkhart, K. L. Maki, M. J. Schertzer, Effects of Interface Velocity, Diffusion Rate, and Radial Velocity on Colloidal Deposition Patterns Left by Evaporating Droplets, J Heat Transfer 139 (2017) 111505 (1-9).